\documentclass[showpacs,preprintnumbers,amsmath,amssymb]{revtex4}
\usepackage{amsmath,amssymb,graphics,epsfig,subfigure}
\usepackage{color}

\begin{document}
\newcommand {\nn} {\nonumber}
\renewcommand{\baselinestretch}{1.3}

\title{Constraining black hole parameters with the precessing jet nozzle of M87*}

\author{Shao-Wen Wei$^{1,2}$ \footnote{weishw@lzu.edu.cn, corresponding author},
Yuan-Chuan Zou$^{3}$\footnote{zouyc@hust.edu.cn},
Yu-Peng Zhang$^{1,2}$\footnote{zyp@lzu.edu.cn},
        Yu-Xiao Liu$^{1,2}$\footnote{liuyx@lzu.edu.cn}}

\affiliation{$^{1}$Lanzhou Center for Theoretical Physics, Key Laboratory of Theoretical Physics of Gansu Province, and Key Laboratory of Quantum Theory and Applications of MoE, Lanzhou University, Lanzhou, Gansu 730000, China,\\
$^{2}$Institute of Theoretical Physics $\&$ Research Center of Gravitation, Lanzhou University, Lanzhou 730000, People's Republic of China,\\
$^{3}$School of Physics, Huazhong University of Science and Technology, Wuhan, 430074, People's Republic of China}

\begin{abstract}
Recently, Cui et al. [Nature \textbf{621}, 711 (2023)] reported that the jet nozzle of M87* exhibits a precession with a period of approximately 11 years. This finding strongly suggests that the supermassive black hole in the core of M87 galaxy is a spinning black hole with a tilted accretion disk. In this paper, our aim is to utilize these observations to preliminarily constrain the parameters of the black hole by using the characteristics of the geodesic motion. Firstly, we investigate the properties of the spherical orbits and the innermost stable spherical orbits with constant radius. The corresponding angular momentum, energy, and Carter constant for both prograde and retrograde orbits are calculated. We find that, compared to equatorial circular orbits, these quantities exhibit significant differences for fixed tilt angles. Moreover, the Carter constant takes positive values for nonvanishing tilt angles. Notably, the presence of misalignment of the orbit angular momentum and black hole spin leads to a precession effect in these spherical orbits. We then make use of these spherical orbits to model the warp radius of the tilted accretion disk, which allows us to determine the corresponding precession period through the motion of massive particles. Further comparing with the observation of M87*, the relationship between the black hole spin and the warp radius is given, through which if one of them is tested, the other one will be effectively determined. Additionally, our study establishes an upper bound on the warp radius of the accretion disk. These findings demonstrate that the precession of the jet nozzle offers a promising approach for testing the physics of strong gravitational regions near a supermassive black holes.
\end{abstract}



\maketitle

\section{Introduction}
\label{secIntroduction}

A few years ago, the Event Horizon Telescope (EHT) Collaboration released the first event-horizon-scale image of the supermassive black hole located at the center of the M87 Galaxy \cite{Akiyama}. This remarkable achievement unveiled the distinctive ring structure, which represents the characteristic features of the black hole shadow, an extraordinary phenomenon arising from strong gravitational effects. Consequently, investigating these shadow patterns provides us with valuable insights into the properties of supermassive black holes. By modeling M87* with the Kerr black hole, the EHT Collaboration demonstrated the remarkable consistency between the observations of the black hole shadow and the predictions of general relativity. Subsequently, EHT Collaboration also unveiled the shadow of the supermassive black hole located at the center of our Milky Way \cite{Akiyamab}. These breakthrough observations have sparked significant interest in the study of black hole shadows, with the objective of constraining the parameters of these celestial objects \cite{Kocherlakota}. However, due to the precision required in capturing these images, this study remains a huge challenge.

On the other hand, the observation of M87* presents a distinctive opportunity to explore the relationship between the relativistic jet and the black hole accretion disk. In a recent study \cite{Lu}, the high-resolution imaging revealed a ring-like structure with a diameter of approximately $8.4$ Schwarzschild radii at a wavelength of 3.5 mm. This remarkable phenomenon further signifies the connection between the jet and the accretion flow surrounding M87*.

Very recently, basing on radio observations spanning 22 years, Cui et al. in Ref. \cite{Yuzhu} reported a remarkable finding that there is a periodic variation in the position angle of the jet with a period of approximately 11 years. By extensive general relativistic magnetohydrodynamics simulation of the settings closely resembling the M87* system, it was suggested the disk and jet precess in a tilted disk. Furthermore, this observation can be attributed to the Lense-Thirring precession resulting from the misalignment of the orbital angular momentum and the black hole spin. Consequently, it strongly indicates that the supermassive black hole in the core of M87 galaxy is a spinning black hole with a tilted accretion disk deviating from the equatorial plane. The half-opening angle of the precession cone is estimated to be $1.25\pm0.18$ degrees, while the angular velocity of precession is measured to be $0.56\pm0.02$ radians per year, leading to a precession period of $11.24\pm0.47$ years.

\begin{figure}
\center{\includegraphics[width=5cm]{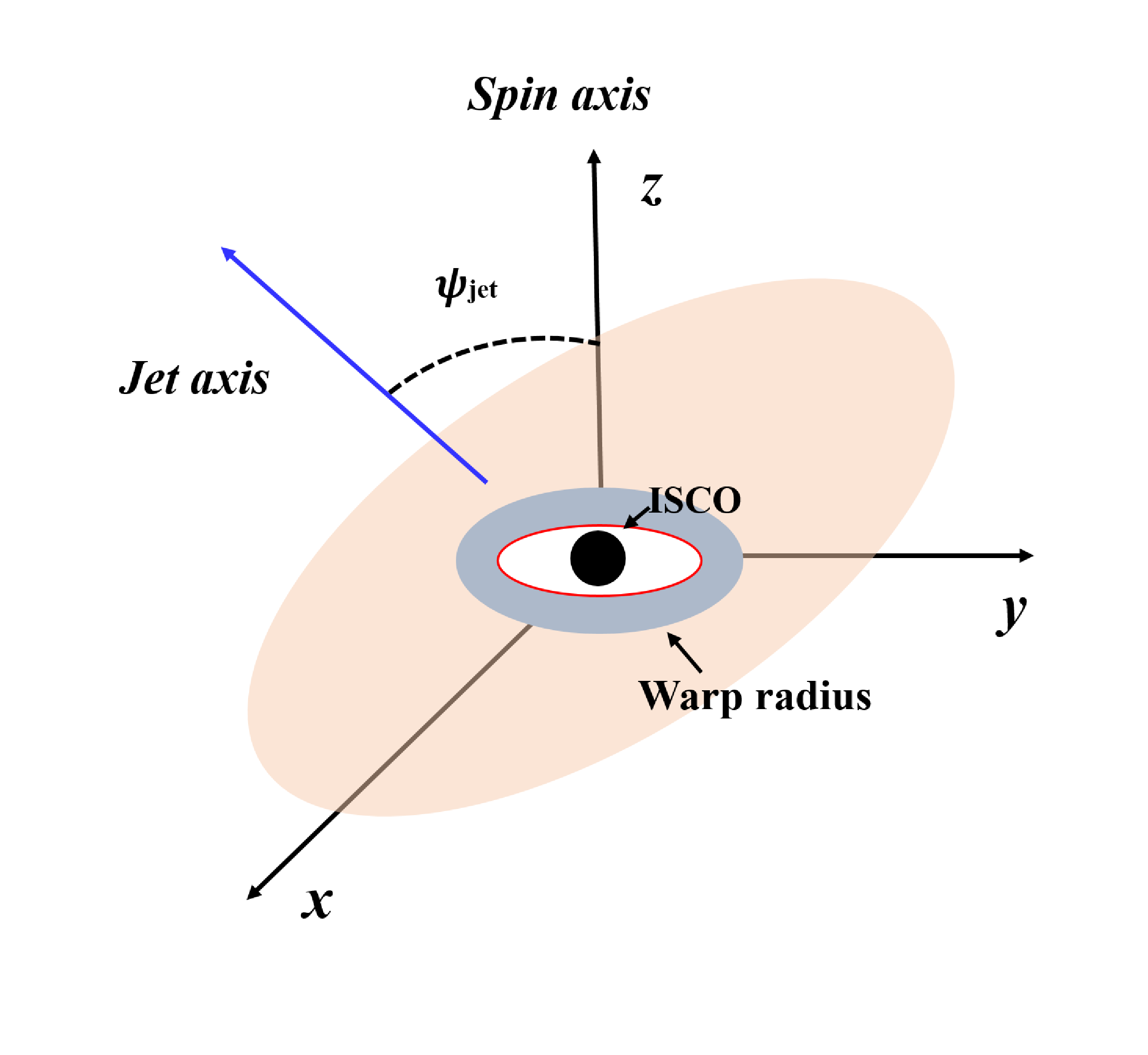}}
\caption{Sketch figure of the tilted accretion disk (pink color), jet axis (blue color), and black hole spin axis ($z$-axis). Near the ISCO, the disk is pulled back to the equatorial plane described by the $xy$ plane.}\label{pSketchFigure}
\end{figure}

The study of tilted accretion disks has been extensively explored in previous works \cite{Petterson, Ostriker}. It is also related with quasi-periodic oscillations \cite{Fragile}. For clarity, we provide a schematic representation of the tilted disk in Fig. \ref{pSketchFigure}. The simulation figure can also be found in Ref. \cite{Lodato}. The figure clearly illustrates the presence of an angle $\psi_{jet}$ that quantifies the misalignment between the orbital angular momentum and the black hole spin. At larger radial distances, the accretion disk, depicted in color, exhibits a tilt angle $\psi_{jet}$. As the radial distance decreases, the gravitational force of the black hole causes the tilt angle of the disk to decrease. Upon reaching a characteristic radius known as the warp radius, the disk is pulled back into the equatorial plane. Furthermore, as particles exceed the equatorial innermost stable circular orbit (ISCO), they shall undergo a rapid fall into the black hole, and disappear behind the event horizon.

For the tilted accretion disks or the motion of a star orbiting the supermassive black hole, the precession induced by the Lense-Thirring effects is supposed to be a common phenomenon that can be traced back to early work. For examples, the possible jet precession and periodic modulation of disk luminosity are observed in the Swift J164449.3+573451 flare \cite{Stone}. The X-ray quasi-periodic eruptions of GSN069 showed an alternative short and long recurring time, which can also be well explained by the orbital precession of the star near the supermassive black hole \cite{Xian}. By making use of the approximation of the geodesic, the Lense-Thirring precession of the trajectory for a star orbiting supermassive rotating black hole was considered in Refs. \cite{Vokrouhlicky,Karasd}. It is also required to explain precessing radio jet of OJ287 on a time-scale of about 22 yr \cite{Britzen}.

In this paper, we aim to present a toy model for quick and effective estimation of the black hole parameters by utilizing the precession period, which might provide a preliminary explanation of the observation and generate interest for further research. For this purpose, we utilize geodesic motion, while making the following assumptions. Firstly, we assume that the motion of the disk particles at each radial distance can be accurately described by the spherical orbits with constant radius while deviating from the equatorial plane \cite{Wilkins,Goldstein,Dymnikova,Shakura,Teo,Rana,Kopacek}. This assumption is well-justified as these orbits tend to be circularized due to the friction within the disk. Secondly, we assume that the jet originates near the warp radius and is oriented perpendicular to the accretion disk. It is important to note that the warp radius is larger than the radius of the ISCO, and thus may not necessarily coincide with it. Lastly, as a third assumption, we consider the precession axis provided in Ref. \cite{Yuzhu} as the axis of the black hole spin.

At first, we study the spherical orbits and innermost stable spherical orbits (ISSOs) through the motion of a test massive particle in the Kerr black hole background. The corresponding radius, angular momentum, energy, and the Carter constant are calculated. Their behaviors with the tilt angle is also examined in detail. Then by making use of the equation of motion of the particle, the angular velocity of the precession, as well as the period,  for a spherical orbit are obtained. Finally, modeling the M87* with the Kerr black hole, we constrain the black hole parameter and the warp radius via the observed period of the precession.

The present study is structured as follows. In Sec. \ref{dddd}, we devote our study to the analysis of spherical orbits and ISSOs. Based on these results, Sec. \ref{ptco} focuses on the calculation of precession for these orbits. In Section \ref{m87s}, we employ the observed period of the precession to constrain the black hole spin and warp radius. Finally, our results are summarized and discussed in Section \ref{Conclusion}.

\section{Spherical orbits}
\label{dddd}

Spherical orbits are the orbits with constant radius. If the trajectory is limited in the equatorial plane, it will be the conventional circular orbit. Otherwise, it is spherical orbit which deviates from the equatorial plane. The original study can be traced back to Wilkins \cite{Wilkins} for the extremal Kerr black hole. The numerically results can be found in Refs. \cite{Goldstein,Dymnikova}. The parameterization forms of the energy and angular momentum are given in Refs. \cite{Shakura,Teo}. It is well known that, along each spherical orbit, the energy, angular momentum, and the Carter constant are preserved. More recently, the numerical results of these constants were presented in Ref. \cite{Kopacek}.

Before discussing the precession of the spherical orbits, in this section, we would like to investigate the radius and constants of the motion with and without inclination to the equatorial plane. The potential properties are also wished to be uncovered.

Let us start with the Kerr black hole. In the Boyer-Linquist coordinates, the Kerr black hole can be described by the following line element
\begin{eqnarray}
 ds^{2}=-\frac{\Delta}{\rho^{2}}\bigg(dt-a\sin^{2}\theta d\phi\bigg)^{2}
        +\frac{\rho^{2}}{\Delta}dr^{2}+\rho^{2}d\theta^{2}\nonumber\\
        +\frac{\sin^{2}\theta}{\rho^{2}}\bigg(adt-(r^{2}+a^{2}) d\phi\bigg)^{2},
\end{eqnarray}
where the metric functions are
\begin{eqnarray}
 \rho^{2}&=&r^{2}+a^{2}\cos^{2}\theta,\\
 \Delta&=&r^{2}-2Mr+a^{2}.
\end{eqnarray}
Solving $\Delta=0$, it is easy to obtain the radii of the black hole horizons
\begin{eqnarray}
 r_{\pm}=M\pm\sqrt{M^2-a^2}.
\end{eqnarray}
Obviously, for a black hole with horizon, its spin must be $a/M\in(0, 1)$.

Using the Hamilton-Jacobi method, the geodesics of a massive particle around the spinning Kerr black hole takes the following forms
\begin{eqnarray}
 \rho^{2}\frac{dr}{d\tau}&=&\pm\sqrt{\mathcal{R}(r)},\label{radia}\quad\\
 \rho^{2}\frac{d\theta}{d\tau}&=&\pm\sqrt{\Theta(\theta)},\label{tth}\\
 \rho^{2}\frac{d\phi}{d\tau}&=&-\frac{P_{\theta}}{\sin^2\theta}+\frac{aP_{r}}{\Delta},\\
 \rho^{2}\frac{dt}{d\tau}&=&-aP_{\theta}+\frac{(r^2+a^2)P_{r}}{\Delta},\label{dtt}
\end{eqnarray}
where $\tau$ is the affine parameter along the geodesics, and
\begin{eqnarray}
 P_{\theta}&=&aE\sin^2\theta-l,\\
 P_r&=&E(r^2+a^2)-al,\\
 \mathcal{R}&=&P^2_{r}-\Delta(r^2+Q+(aE-l)^2),\\
 \Theta&=&Q+(aE-l)^2-a^2\cos^2\theta-\frac{P_{\theta}^2}{\sin^2\theta}.\label{ttha}
\end{eqnarray}
The symbols $E$ and $l$ represent the energy and angular momentum per unit mass of the test particle, respectively, and are associated with the Killing fields $\partial_t$ and $\partial_\phi$. The Carter constant $Q$, which corresponds to the Killing-Yano tensor, is another conserved quantity along each geodesic.

For an orbit confined to the equatorial plane, we have $\theta=\frac{\pi}{2}$ and $\frac{d\theta}{d\tau}=0$, resulting in a vanishing Carter constant $Q=0$. Consequently, for the orbits deviating from the equatorial plane, the Carter constant does not vanish. However, it remains constant along the trajectory of a massive particle. Let us now focus on a geodesic in the $\theta$-motion. By examining Eqs. (\ref{tth}) and (\ref{ttha}), it is apparent that the $\theta$-motion of the massive particle exhibits symmetry about $\theta=\frac{\pi}{2}$. Therefore, for a bounded motion, the value of $\theta$ must be confined within the range [$\frac{\pi}{2}-\zeta$, $\frac{\pi}{2}+\zeta$], where $\zeta\in[0, \frac{\pi}{2}]$ represents the maximum half-opening angle of the motion along the $\theta$ direction. For convenience, we refer to $\zeta$ as the tilt angle. Considering the particle turns back at this point, we have $\frac{d\theta}{d\tau}=\Theta=0$, which gives
\begin{eqnarray}
 Q=l^2 \tan^2\zeta+a^2 \left(1-E^2\right)\sin^2\zeta.\label{cca}
\end{eqnarray}
For a geodesic with bounded $\theta$-motion, the quantities $l$ and $E$ remain constant. Additionally, if the maximum half-opening angle $\zeta$ is given as a priority, the Carter constant will also be determined. Subsequently, we can analyze the $r$-motion of the massive particle. In the following discussion, we specifically focus on these orbits with constant radius. On the other hand, according to equation (\ref{cca}), it can be observed that the Carter constant vanishes for the equatorial motion with $\zeta=0$, while it takes positive values for these orbits off the equatorial plane with bounded energy $E<1$. Moreover, negative values of the Carter constant indicate
\begin{eqnarray}
 E>\sqrt{1+\frac{l^2}{a^2\cos^2\zeta}}.
\end{eqnarray}
Obviously, the energy $E>1$.

\subsection{Spherical orbits}

Quite recently, these constants along the ISSOs, marginally bound spherical orbits, and the innermost stable polar orbit were calculated in Ref. \cite{Kopacek}. It was also demonstrated in Ref. \cite{Zahrani} that, the precession of these spherical orbits was obtained from the perspective of a local observer. For the purpose of calculating the period of the precession for a distant observer, we aim to numerically obtain the energy, angular momentum, and Carter constant for a general spherical orbit, and compare the result with that of Ref. \cite{Kopacek} for uncovering more sub detail of these spherical orbits and ISSOs. Furthermore, we will examine the stability of these orbits.

From Eq. (\ref{radia}), the radial $r$-motion can be expressed as
\begin{eqnarray}
 \left(\rho^2\frac{dr}{d\tau}\right)^2+V_{eff}=0,
\end{eqnarray}
where the effective potential reads
\begin{eqnarray}
 V_{eff}=-\mathcal{R}(r).
\end{eqnarray}
Note that $\mathcal{R}(r)$ depends on the Carter constant, which has a close relation with the tilt angle $\zeta$. Therefore, it is actually dependent of $\zeta$. The constant radius orbits require
\begin{eqnarray}
 \mathcal{R}=\mathcal{R}'=0,\label{ccc}
\end{eqnarray}
where the prime symbol denotes the derivative with respect to $r$. From above equation, we can determine the energy and angular momentum corresponding to an orbit with a given radius $r$. The explicit form of these quantities is omitted for brevity. Additionally, the stability of these orbits is determined by the value of $\mathcal{R}''$. Specifically, a negative value of $\mathcal{R}''$ corresponds to a stable orbit, while a positive value indicates an unstable one.

\begin{figure}
\center{\subfigure[]{\label{Veffa}
\includegraphics[width=5cm]{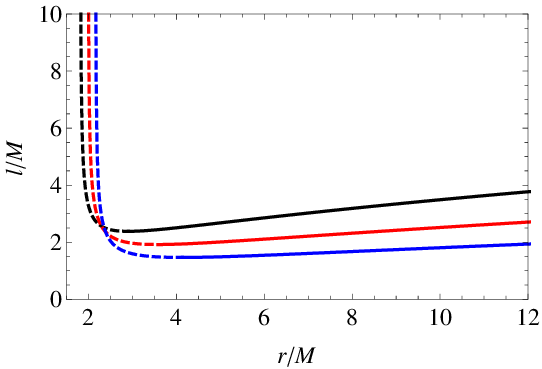}}
\subfigure[]{\label{Veffb}
\includegraphics[width=5cm]{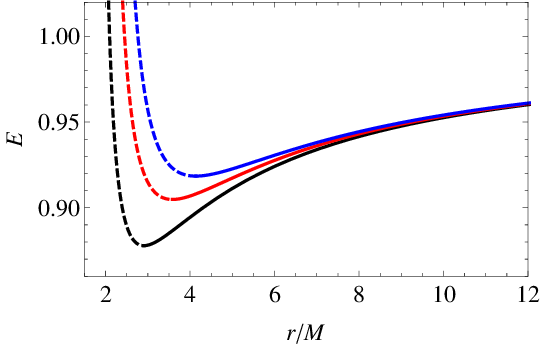}}
\subfigure[]{\label{Veffc}
\includegraphics[width=5cm]{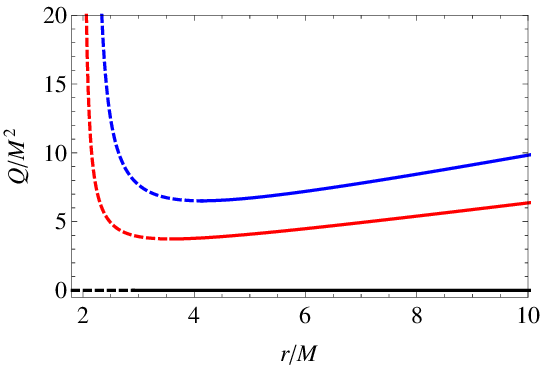}}\\
\subfigure[]{\label{Veffd}
\includegraphics[width=5cm]{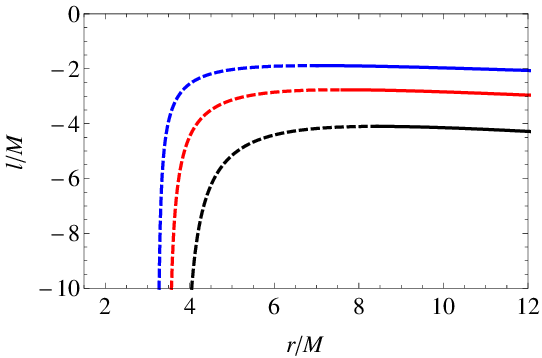}}
\subfigure[]{\label{Veffe}
\includegraphics[width=5cm]{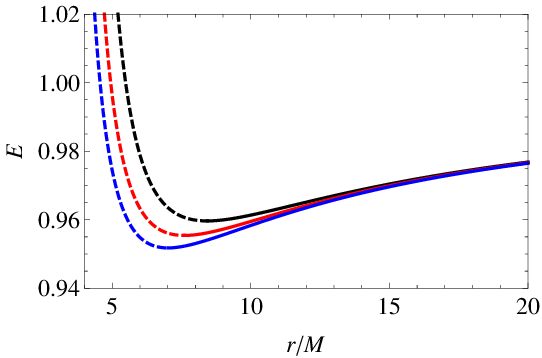}}
\subfigure[]{\label{Vefff}
\includegraphics[width=5cm]{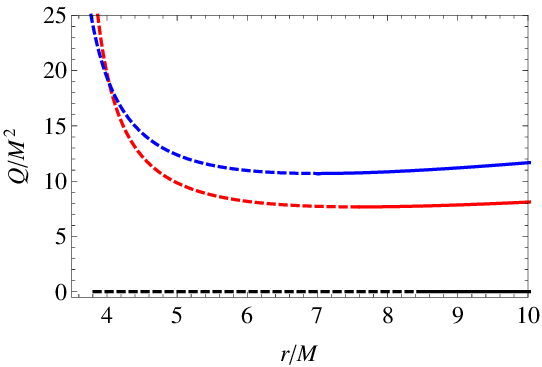}}}
\caption{Angular momentum, energy, and Carter constant for the spherical orbits with $a$=0.8, and $\zeta$=0, $\frac{\pi}{4}$, and $\frac{\pi}{3}$. Upper row is for the prograde orbits and lower row is for the retrograde orbits. (a) $l/M-r/M$. (b) $E-r/M$. (c) $Q/M^2-r/M$. (d) $l/M-r/M$. (e) $E-r/M$. (f) $Q/M^2-r/M$. $\zeta$=0, $\frac{\pi}{4}$, and $\frac{\pi}{3}$ from top to bottom for (a) and (e), and from bottom to top for other figures. Solid and dashed curves are for the stable and unstable spherical orbits, respectively.}\label{ppVefff}
\end{figure}

We present the variations of the angular momentum, energy, and Carter constant as a function of the radius of the spherical orbits in Fig. \ref{ppVefff}. Notably, both the angular momentum and energy exhibit non-monotonic behaviors, as illustrated in Figs. \ref{Veffa}, \ref{Veffb}, \ref{Veffd}, and \ref{Veffe}. Taking the stability into account, we represent stable spherical orbits with solid curves and unstable spherical orbits with dashed curves. It is worth noting that these two types of orbits are connected by the ISSOs. The unstable spherical orbits possess a minimum radius, leading the energy and angular momentum to diverge towards positive infinity. In contrast, the stable spherical orbits originate from the ISSO and extend towards radial infinity. It is worth noting that the energy of the stable spherical orbits is always bounded below 1. Moreover, as the tilt angle $\zeta$ increases, the absolute value of the angular momentum $l$ decreases, and the energy decreases for retrograde orbits but increases for prograde orbits. Notably, as illustrated in Figs. \ref{Veffc} and \ref{Vefff}, the Carter constant $Q$ always vanishes for $\zeta=0$, while it takes positive values for non-zero $\zeta$. In the case of unstable spherical orbits, $Q$ decreases as the radius increases, whereas it increases for stable spherical orbits. These findings provide novel insights into the behavior of the Carter constant for the spherical orbits.

\begin{figure}
\center{\subfigure[]{\label{lzeta}
\includegraphics[width=5cm]{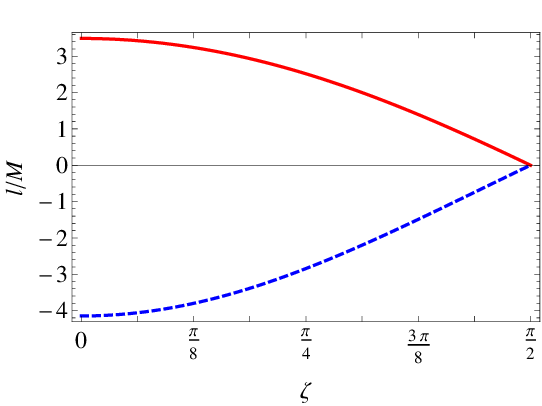}}
\subfigure[]{\label{Ezeta}
\includegraphics[width=5cm]{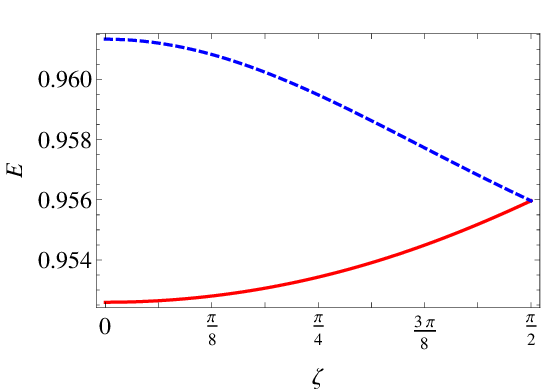}}
\subfigure[]{\label{Qzeta}
\includegraphics[width=5cm]{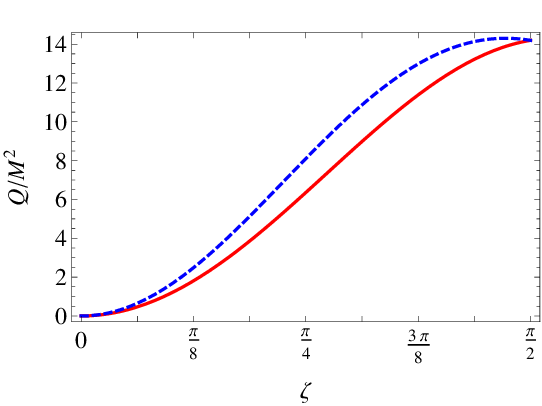}}}
\caption{Angular momentum, energy, and Carter constant as a function of the tilt angle $\zeta$ with $a/M=0.8$ and $r/M=10$. (a) $l/M-\zeta$. (b) $E-\zeta$. (c) $Q/M^2-\zeta$. The red solid curves and blue dashed curves are for the prograde and retrograde spherical orbits.}\label{ppQzeta}
\end{figure}

To illustrate the behaviors of the angular momentum, energy, and Carter constant as a function of the tilt angle, we present them in Fig. \ref{ppQzeta} for the specific case of $a/M=0.8$ and $r/M=10$, where only the stable spherical orbits exist. From the figure, it is evident that the angular momentum $l/M$ decreases with increasing tilt angle $\zeta$ for prograde spherical orbits, while it increases for retrograde spherical orbits. The opposite trend is observed for the energy $E$, where it increases with $\zeta$ for prograde orbits and decreases for retrograde orbits. Notably, although the energy and angular momentum exhibit different values at $\zeta=0$, they tend to converge to the same values at $\zeta=\frac{\pi}{2}$. The results depicted in Fig. \ref{Qzeta} demonstrate that the Carter constant vanishes for both prograde and retrograde spherical orbits when the tilt angle $\zeta$ is zero. Subsequently, $Q/M^2$ exhibits rapid increase for retrograde orbits. Notably, the Carter constant tends to converge to the same value as $\zeta$ approaches $\frac{\pi}{2}$. A notable characteristic of the retrograde orbits is the non-monotonic behavior of the Carter constant $Q/M^2$. Detailed calculations reveal that it decreases within the range $\zeta \in (1.48, \frac{\pi}{2})$, reaching its maximum value of $Q/M^2=14.30$ at $\zeta=1.48$. Similar behaviors can also be observed by varying the spin of the black hole.

\subsection{Innermost stable spherical orbits}

As demonstrated earlier, the ISSO serves as a connection between the stable and unstable spherical orbits. In this subsection, our attention will be directed towards these special orbits characterized by a non-zero tilt angle $\zeta$.

From the effective potential, the ISSO can be determined by
\begin{eqnarray}
 \mathcal{R}=\mathcal{R}'=\mathcal{R}''=0.\label{isc}
\end{eqnarray}
On the other hand, we can find from Fig. \ref{ppVefff} that ISSOs always locate at the extremal points of each curve of the spherical orbits. Therefore, we can solve one of the following equations
\begin{eqnarray}
 \left(\frac{dl}{dr}\right)_{a,\zeta}=0, \quad  \left(\frac{dE}{dr}\right)_{a,\zeta}=0, \quad\left(\frac{dQ}{dr}\right)_{a,\zeta}=0,
\end{eqnarray}
for the ISSO. Note that these quantities, angular momentum $l$, energy $E$, and Carter constant $Q$, are related with the spherical radius $r$.

\begin{figure}
\center{\subfigure[]{\label{RISCO}
\includegraphics[width=5cm]{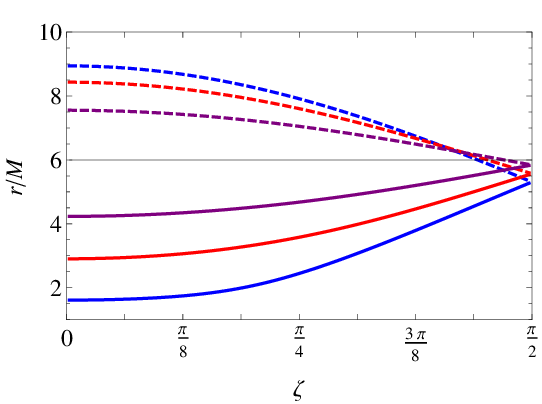}}
\subfigure[]{\label{LISCO}
\includegraphics[width=5cm]{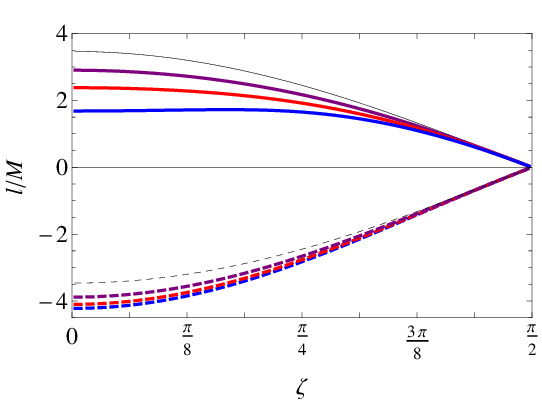}}\\
\subfigure[]{\label{EISCO}
\includegraphics[width=5cm]{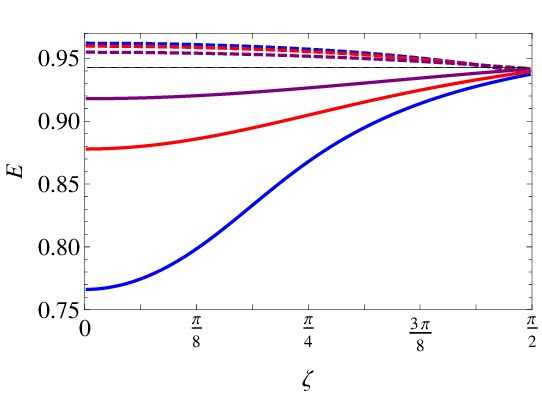}}
\subfigure[]{\label{QOISCO}
\includegraphics[width=5cm]{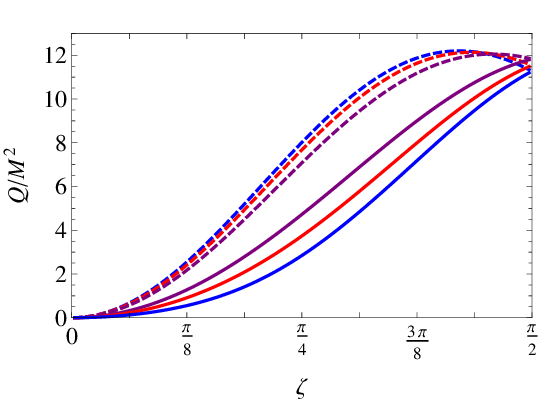}}}
\caption{Characteristic quantities of the ISSO as a function of the tilted angular $\zeta$. (a) $r/M-\zeta$. (b) $l/M-\zeta$. (c) $E-\zeta$. (d) $Q/M^2-\zeta$. Black hole spin $a/M$=-0.98, -0.8, -0.5, 0.5, 0.8, 0.98 for these thick curves from top to bottom in (a), (c), (d), and from bottom to top in (b). The solid thin curves are for the case with $a/M=0$. Dashed and solid curves are for the retrograde and prograde ISSOs.}\label{ppQOISCO}
\end{figure}

By considering various values of the black hole spin, we present the radius, angular momentum, energy, and Carter constant in Fig. \ref{ppQOISCO} by numerically solving the conditions (\ref{isc}). Dashed and solid curves correspond to prograde and retrograde orbits, respectively. It is evident that the angular momentum, energy, and Carter constant exhibit similar behaviors to those observed in the stable spherical orbit depicted in Fig. \ref{ppQzeta}. From the observations in Fig. \ref{RISCO}, it is evident that the radius of the prograde ISSO is consistently smaller than 6$M$ and increases as the tilt angle $\zeta$ increases. Conversely, for retrograde ISSOs with a non-zero black hole spin, their radii decrease with the increasing of $\zeta$. This is exactly consistent with the result of Fig. 5 of Ref. \cite{Kopacek}. Since the energy and angular momentum of the ISSOs are not respectively given, we will not discussed them here. More importantly, when $\zeta$ reaches certain critical value, the radius of the retrograde ISSO becomes smaller than 6$M$, which corresponds to the scenario of a Schwarzschild black hole. This result has not been observed in Ref. \cite{Kopacek}, and provides us with a substructure of the ISSOs. It is natural to conjecture that this behavior may be linked to the non-monotonic behavior of the Carter constant, as illustrated in Fig. \ref{QOISCO}. By taking $\zeta=\pi/2$ in Fig. \ref{QOISCO}, we find the Carter constant decreases with the black hole spin. This confirms the result given in Ref. \cite{Kopacek} for the innermost stable polar orbit. Obviously, this pattern of the radius differs significantly from that of the equatorial ISCOs.

On the other hand, the dependency of the orbits is also worth studying. From Fig. 4 of Ref. \cite{Kopacek}, one observes that for different Carter constant corresponding to the tilt angle, the radius of ISSOs coincides at 6$M$ for $a$=0. This indicates the dependency. However, for $a$=0, the black hole is exactly the Schwarzschild one possessing an ISCO at 6$M$. So this is induced by the spherical symmetry of black hole. From Fig. \ref{RISCO}, we observe that there may exist a dependency of the ISSOs at certain radius for the retrograde case with different black hole spin. After a detailed numerical calculation, we find these dashed curves coincide at $r/M$=6.23, 6.28, and 6.43, respectively. This indicates that there exists no such dependency for the ISSOs.

\section{Precession of spherical orbits}
\label{ptco}

In the case of equatorial circular orbits, the angular momentum is aligned parallel to the black hole spin, resulting in the orbit being confined to the initial equatorial plane. However, when the orbits become tilted, an angle is formed between the directions of the orbital angular momentum and the black hole spin. This inclination gives rise to the precession, commonly referred to as Lense-Thirring precession, causing the orbit plane to deviate from its initial plane. In this section, our focus will be on investigating the precession of these spherical orbits.

\begin{figure}
\center{\subfigure[]{\label{GuijiQ}
\includegraphics[width=6cm]{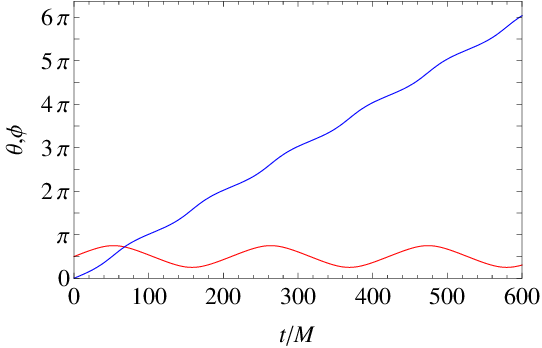}}
\subfigure[]{\label{GuijiT}
\includegraphics[width=6cm]{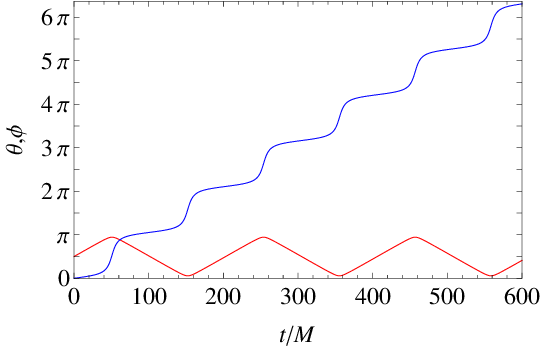}}}
\caption{Evolutions of the angular coordinates $\theta$ (bottom red curves) and $\phi$ (top blue curves) with $r/M=10$ and $a/M=0.9$. (a) $\zeta=\frac{\pi}{4}$. (b) $\zeta=\frac{4\pi}{9}$.}\label{ppGuijiT}
\end{figure}

To provide a clear visualization, we present the $\theta$ and $\phi$ motions for $a/M=0.9$ and $r/M=10$ in Fig. \ref{ppGuijiT}. In Fig. \ref{GuijiQ}, we set the tilt angle as $\zeta=\frac{\pi}{4}$. It is obvious that the $\theta$ motion is confined within finite values, while the $\phi$ motion increases boundlessly with the coordinate time $t$. This behavior approximates a linear trend with slight deviations, indicating that the particle maintains a constant velocity in the $\phi$ direction. Furthermore, when considering a larger tilt angle, such as $\frac{4\pi}{9}$, the motion characteristics are displayed in Fig. \ref{GuijiT}. The $\theta$ motion remains bounded over time, but with notable difference. It exhibits a step-like behavior, suggesting that the particle experiences an abrupt change in its trajectory. The slope of the curve represents the velocity in the $\phi$ direction, indicating that the particle attains higher velocities near the north and south poles while having lower velocities near the equatorial plane.

The precession angular velocity $\omega_{t}$ of the spherical orbit can be calculated with
\begin{eqnarray}
 \omega_{t}=\frac{\Delta\phi-2\pi}{T_{\theta}},
\end{eqnarray}
where $T_{\theta}$ represents the period of the $\theta$ motion, while $\Delta\phi$ quantifies the difference in angular $\phi$ over one period of the $\theta$ motion. It is important to note that the $\omega_{t}$ defined here is for the distant observer, as opposed to the local observer defined in Ref. \cite{Zahrani}. To extract the precession information from the motions depicted in Fig. \ref{ppGuijiT}, we count the coordinate time and the corresponding $\phi$ when the particle crosses three successive maximum values of $\theta$, as presented in Table I. Utilizing this data, we can calculate the precession angular velocity $\omega_{t}$ for each tilted spherical orbit. For the tilted orbit with $\zeta=\frac{\pi}{4}$, we obtain $\omega_{t}=0.00151$ and $0.00151$ for the two consecutive periodic motions in $\theta$. Similarly, for the orbit with $\zeta=\frac{4\pi}{9}$, we find $\omega_{t}=0.00171$ and $0.00170$. These results also indicate that the calculation error is well controlled for our calculations.

\begin{table}[]
\setlength{\tabcolsep}{2.5mm}{\begin{tabular}{ccccc}\hline\hline
  &    & $\theta_{max}^{1}$  & $\theta_{max}^{2}$ &  $\theta_{max}^{3}$ \\\hline\hline
$\zeta=\frac{\pi}{4}$  &  $t$      & 52.63957  & 263.35951 & 474.06870 \\
                                    & $\phi$   & 1.64887    & 8.25074     & 14.85215 \\\hline
$\zeta=\frac{4\pi}{9}$&  $t$      & 50.73793  & 253.81609 & 456.87905 \\
                                    & $\phi$   & 1.65344    & 8.28467     & 14.91318 \\\hline\hline
\end{tabular}
\caption{Values of $t$ and $\phi$ when the particle crosses the maximal $\theta$ for these two spherical orbits described in Fig. \ref{ppGuijiT}.}\label{tab}}
\end{table}

\section{Constrains of M87*}
\label{m87s}

In Ref. \cite{Yuzhu}, Cui et al. reported a notable precession of the jet axis, with an observed period of approximately 11 years. This observation provides an opportunity to constrain the parameters of the black hole as expected. It is important to note that the sign of the precession angular velocity cannot be determined definitively. Therefore, in this section, one needs to consider both the prograde and retrograde cases to account for the possible scenarios.

In general, the precession of a jet or an accretion disk is influenced by various factors, including the properties of the surrounding environment, such as the accretion flow and the presence of external bodies, which introduce additional forces or torques. Thus, the precession frequency being completely decoupled from the influence of the outer accretion flow and external bodies would constitute an exceptional case rather than the expected scenario in most astrophysical systems. However, for a simplified or toy model, it is beneficial to assume idealized conditions where the precession frequency is decoupled from the influence of the outer accretion flow and external bodies, for the sake of simplicity and initial estimation. Such a model can serve as a starting point for preliminary analysis and feasibility assessment, enabling quick and effective estimations before more detailed and complex simulations or analysis are feasible.

To match the observable, we define the period of the precession for the spherical orbit as $T=\frac{2\pi}{\omega_{t}}$, or
\begin{eqnarray}
 T&=&\frac{2\pi}{\omega_{t}}\frac{GM_{\odot}}{c^{3}}\left(\frac{M}{M_{\odot}}\right)\nonumber\\
   &\approx&6.394\times 10^{-3}\times\frac{1}{\omega_{t}}\;\;(year),
\end{eqnarray}
when the unit is restored.

\begin{figure}
\center{\subfigure[]{\label{PeriodRetr}
\includegraphics[width=6cm]{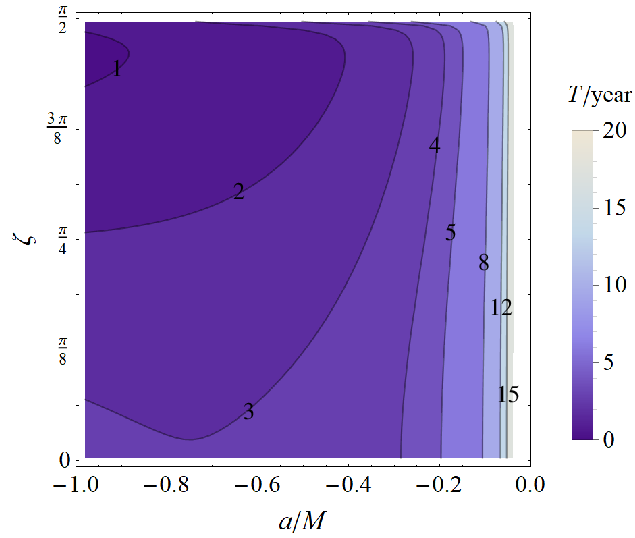}}
\subfigure[]{\label{Periodprog}
\includegraphics[width=6cm]{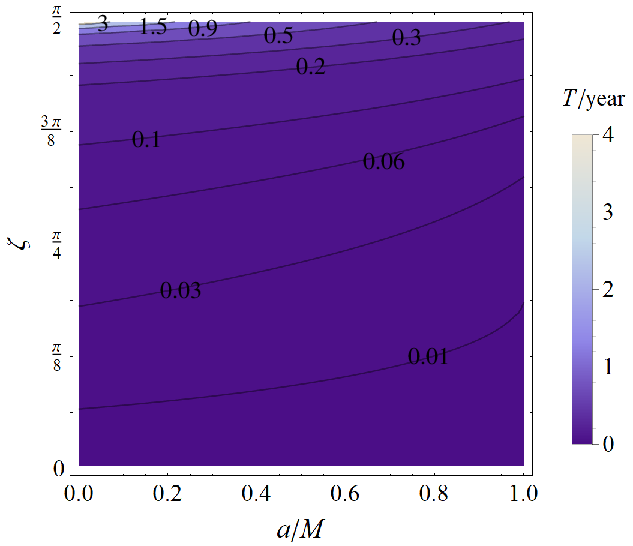}}}
\caption{The period of the precession angular velocity $T$ (in unit of year) for the ISSOs in the $\zeta$-$a/M$ plane. (a) Retrograde ISSO case. (b) Prograde ISSO cases. The period remains predominantly below 5 years across most parameter regions for both retrograde and prograde cases. However, a long period can be reached at extremely small spins for retrograde case, and small spin and large tilt angle $\zeta$ for prograde case. }\label{ppPeriodprog}
\end{figure}

Here, we primarily consider two cases: the jet originating from the inner and outer parts of the accretion flow, respectively modeled by the ISSO and spherical stable orbits. These cases may yield distinct and insightful results.

First, we consider the location of the ISSO as the characteristic radius of the accretion disk. By varying the values of the black hole spin $a/M$ and the tilt angle, we determine the radius, energy, angular momentum, and the Carter constant of the ISSO. Utilizing equations (\ref{radia})-(\ref{dtt}), we calculate the trajectory of the particle. Then, the precession angular velocity $\omega_{t}$ and its corresponding period will be obtained.

Following the above approach, we present the period of the precession angular velocity in Fig. \ref{ppPeriodprog} for the M87* black hole with a mass of $M=6.5\times10^9M_{\odot}$ for the retrograde and prograde orbits, respectively. Examining the $\zeta$-$a/M$ plane, it becomes evident that the period remains predominantly below 5 years across most parameter regions for both retrograde and prograde cases. However, for retrograde ISSOs shown in Fig. \ref{PeriodRetr}, their period can extend to higher values, reaching up to 20 years when the black hole spin is extremely small. This behavior holds true regardless of the specific value of the tilt angle $\zeta$, representing a universal result. In general, the precession angular velocity $\omega_{t}$ exhibits a proportional relationship with the black hole spin, thereby explaining this result in a straightforward manner. For the prograde ISSO case shown in Fig. \ref{Periodprog}, in order to achieved a long period, the result indicates that the black hole spin must be small and the tilt angle $\zeta$ approach $\pi/2$ corresponding to an accretion disk perpendicular to the equatorial plane, which is not feasible for M87*. Therefore, if the ISSO is the characteristic radius, the accretion disk should be retrograde one with extremely small spin, while the tilt angle remains to be fixed.

On the other hand, the observations of the shadow and jet by the EHT suggest that M87* may possess a wide accretion disk, which indicates that it is inappropriate to describe the disk by the ISSO. Fortunately, the study of tilted accretion disks reveals that the ISSO does not represent the characteristic radius of the whole disk \cite{Petterson, Ostriker}. As we move to larger radii, the disk becomes increasingly tilted. However, as the radial distance decreases, the tilt angle of the accretion disk also reduces. Eventually, at the warp radius, the tilt angle vanishes. Moreover, as particles in the disk approach the ISSO, they rapidly plunge into the black hole. During this process, it is nature to think that the jet originates from the warp radius, and which is believed to lie within the range of (6$M$, 20$M$). This case is different from that of Ref. \cite{Yuzhu}, where the jet precession is induced by the precession of the inner hot inclined torus. Taking this into consideration, our aim is to investigate the precession period at different radii of the spherical orbits within the accretion disk and to present the potential observation significance.

\begin{figure}
\center{\subfigure[]{\label{TwMa}
\includegraphics[width=6cm]{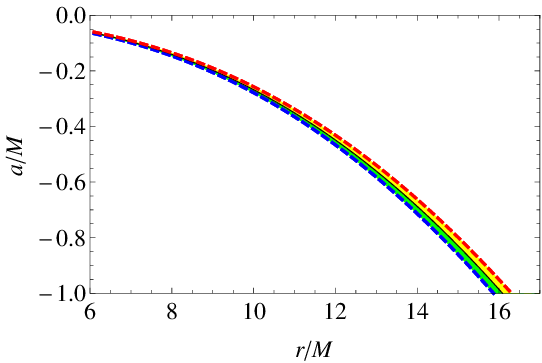}}
\subfigure[]{\label{Twab}
\includegraphics[width=6cm]{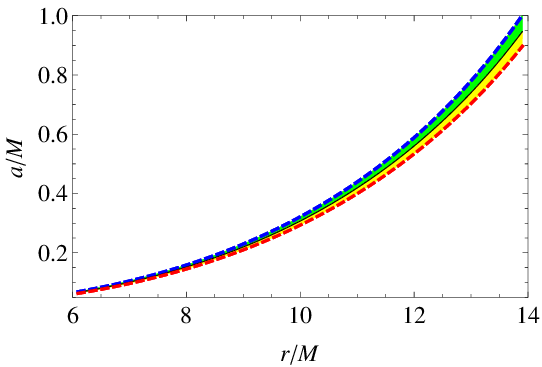}}}
\caption{Constrains of the black hole spin and warp radius of the accretion disk with the precessing period of the jet nozzle of M87*.  The tilt angle $\zeta=1.25^{\circ}$. The solid curve corresponds the period $T$=11.24 years. The  blue and red dashed curves are for $T$=10.77 and 11.71 years.}\label{ppTwab}
\end{figure}

Based on the observations of M87*, the tilt angle is determined to be $\zeta=1.25^{\circ}$ \cite{Yuzhu}. By varying the black hole spin and the radius of the spherical orbits that characterize the accretion disk, we calculate the precession period. The EHT observation reveals that the period corresponds to $11.24\pm0.47$ years. Consequently, in Fig. \ref{ppTwab}, we represent this observed period in the $a/M-r/M$ plane. The solid curve corresponds to a period of $T=11.24$ years. Additionally, the blue and red dashed curves represent the upper and lower bounds of the period, namely $T=10.77$ and $11.71$ years, respectively.

Upon examining the figures, it becomes evident that, for each fixed period, the absolute value of the black hole spin increases with the warp radius. Conversely, for a fixed warp radius, the period decreases as the black hole spin increases. Additionally, for a given black hole spin, the period of the precession increases with the warp radius. Our primary aim is to constrain the black hole spin via the observed precession period. However, the obtained results do not appear to be optimistic. For instance, considering a warp radius of $r=6M$, the black hole spin $a/M$ can be reduced to 0.0604 and 0.0641 for the retrograde and prograde cases, respectively, which is quite close to the case of a nonspinning black hole.

However, an unexpected result emerges at large warp radii. It does not extend to infinity but instead exhibits an upper bound. For instance, in the case of prograde or retrograde tilted accretion disks, we can determine the maximum warp radius
 \begin{eqnarray}
 r_{max}=14M, \quad  16M,
\end{eqnarray}
respectively. Although our model is simplistic and these values should be refined through magnetohydrodynamic simulations where the factors such as the viscosity of the accretion flow are considered, it qualitatively demonstrates the existence of a potential difference between prograde and retrograde scenarios, which may lead to observable effects in the near future. As a result, by utilizing the observed precession period, we can probe the maximum warp radius for the tilted accretion disk. Furthermore, if the observed warp radius aligns within a range of a 2$M$ difference, it becomes feasible to distinguish whether the accretion disk is prograde or retrograde. Consequently, this result holds significant value and warrants further testing through astronomical observations.

On the other hand, we observe that the constraint imposed by the observation becomes tighter for smaller warp radii and lower black hole spins, while it becomes looser for highly spinning black holes and larger warp radii. Despite not obtaining a stringent constraint on the black hole spin, our findings in Fig. \ref{ppTwab} reveal the existence of a constrained region, marked in green and yellow colors, within the parameter space. This strongly indicates that if the warp radius is determined through other astronomical observations, we can subsequently determine the black hole spin. Conversely, the warp radius can also be tested based on the black hole spin. In order to provide convenience for future applications, we provide a high-precision fitting formula that relates the dimensionless black hole spin to the warp radius corresponding to a precession period of $T=11.24$ years
 \begin{eqnarray}
 a/M&=&-0.005742+0.001571(r/M) + 0.000058 (r/M)^2 - 0.000341 (r/M)^3 +0.000006 (r/M)^4,\label{adoda}\\
 a/M&=&0.450882 - 0.212997(r/M) + 0.037536 (r/M)^2 - 0.002674 (r/M)^3+ 0.000091 (r/M) ^4, \label{adodb}
\end{eqnarray}
for retrograde and prograde tilted accretion disks, respectively. These formulas enable us to determine one of the warp radius or black hole spin by knowing the other. This is also a valuable outcome of our simplistic model.

\section{Discussions and conclusions}
\label{Conclusion}

In this paper, we aimed to constrain the parameters of M87* by using the recent observation of the precessing jet nozzle \cite{Yuzhu}. In addition to the shadow, this observation presents another promising approach for testing the properties of the supermassive black holes and investigating the physics within strong gravitational regions.

The presence of a precessing jet axis in M87* suggests that its accretion disk is not in the equatorial plane, but instead exhibits a tilt angle of approximately 1.25$^\circ$ \cite{Yuzhu}. Given that the accretion disk can be effectively described by corresponding spherical orbits, we initiated our investigation by examining the properties of them. These orbits are confined within the $\theta$ motion. As the spherical orbits represents a type of geodesic, the radius, angular momentum, energy, and the Carter constant remain constant along each spherical orbit. Utilizing the equations of motion for massive particles, we obtained the values for both the spherical orbits and the ISSOs. Similar to the equatorial case, the spherical orbits with small radii are found to be radially unstable, while those with larger radii are stable. However, a notable distinction is that the Carter constant no longer vanishes for these spherical orbits, but instead maintains positive values. As expected, the presence of a tilt angle results in the deviations of the angular momentum, energy, and the Carter constant from those of equatorial circular orbits. Our findings reveal that as the tilt angle increases, the angular momentum decreases while the energy increases for the prograde spherical orbits. Conversely, this trend is reversed for the retrograde spherical orbits. Moreover, for both prograde and retrograde spherical orbits, the Carter constant exhibits an increasing behavior from a tilt angle of $\zeta=0$ and reaches a maximum near $\zeta=\frac{\pi}{2}$. Notably, a subtle feature is observed, wherein a slight decrease occurs in the vicinity of $\zeta=\frac{\pi}{2}$ for the retrograde orbits.

The ISCO represents the last stable orbit of the massive particles before they rapidly fall towards the black hole. It is widely accepted as the inner boundary of the accretion disk. In order to provide a more accurate description of the disk, we also examined the behavior of the ISSOs when the tilt angle is non-zero. Our results demonstrate that, for each fixed black hole spin, the angular momentum, energy, and the Carter constant exhibit similar trends as the tilt angle increases. However, the radius of the ISSO exhibits an intriguing behavior. In the case of the equatorial ISCO, the radius for either prograde or retrograde motion is always smaller or larger, respectively, than $6M$, which is the value for a Schwarzschild black hole. However, with an increase in the tilt angle, the retrograde ISSO reveals a notable feature: its radius becomes smaller than $6M$, indicating a distinct characteristic of tilted retrograde ISSOs.

Having obtained the characteristic quantities of the spherical orbits, we proceeded to plot the trajectories for the massive particles moving along these orbits. Due to the presence of the tilt angle, the orbital angular momentum and the black hole spin become misaligned. Consequently, the particle does not return to its initial location after completing one loop, but exhibits a discernible deviation. For instance, following one complete circle in the $\theta$ motion, the particle's $\phi$ angle will experience a small shift. Exploiting this observation, we defined the precession angular velocity $\omega_{t}$ as a measurement of the variation in the orbital plane. Consequently, the precession manifests as a periodic phenomenon with a period denoted as $T$, which precisely corresponds to the period observed in the precessing jet nozzle \cite{Yuzhu}.

Subsequently, we assumed that the jet originates from the location defined by the ISSO and proceeded to calculate the precession period for various black hole spins and tilt angles. The results reveal that, in the majority of the parameter space, the precession period is below five years. However, for extremely slowly spinning black holes, an 11 year period can be achieved. Consequently, based on this pattern, if M87* possesses spin, it must be exceedingly small. Meanwhile, the accretion disk is retrograde with the black hole spin.

On the other hand, our study of the tilted accretion disk reveals that the ISSO may be not the location where the jet originates. Instead, at larger radii, the disk exhibits a significant tilt, while with decreasing radius, the degree of tilt decreases. Remarkably, when the radius exceeds a certain threshold known as the warp radius, the gravitational forces pull the disk back towards the equatorial plane. Ultimately, at the equatorial ISCO, the accretion disk ceases to extend further. Consequently, it is natural to consider the warp radius as the origin of the jet, which is estimated to be within the range of 6$M$ to 20$M$.

Under the assumption mentioned above, we revisited the precession period by setting $\zeta=1.25^{\circ}$. The numerical results indicate that, with a warp radius at $6M$ and an expected period of $T=11.24$ years, the possible black hole spin can be as low as 0.06. At first glance, it may seem challenging to establish a precise constraint on the black hole spin. However, this case might change for a large warp radius. We provide an explicit relationship between the black hole spin and the warp radius, as illustrated in Fig. \ref{ppTwab}. Furthermore, we derived the fitting formulas in (\ref{adoda}) and (\ref{adodb}). Therefore, if we can determine the warp radius through other observations, we can precisely determine the black hole spin using such precession measurements. It is worth noting that the reverse approach is also feasible.

On the other hand, while it may be challenging to constrain the black hole spin precisely, it does impose an upper bound on the warp radius. In the case of prograde and retrograde accretion disks, we find upper bounds of $r/M=$ 14 and 16, respectively. Notably, the difference between these bounds can also serve as a potential distinguishing feature between the prograde and retrograde accretion disks.

In conclusion, while our treatments have notable limitations when considering the complex astrophysical environment and accretion disk fluid model, they still yield insightful results that could spark further interest. For instance, if the jet originates from the inner regions of the accretion flow, our results indicate that a retrograde disk is more feasible and the black hole spin is extremely small. Conversely, if the jet emanates from the outer regions of the accretion flow, the warp radius should have an upper bound, even with a value that needs to be refined by accounting for realistic large-scale flows, magnetohydrodynamics, and radiation effects. Furthermore, it presents an opportunity to test physics, such as hidden dimensions \cite{Banerjee}, in strong gravitational regions proximate to supermassive black holes.

\section*{Acknowledgements}
We warmly thank Prof. Weihua Lei for useful discussions. This work was supported by the National Natural Science Foundation of China (Grants No. 12075103, No. 11875151, No.12105126, and No. 12247101) and the Major Science and Technology Projects of Gansu Province.

\end{document}